\documentclass[aps,pra,twocolumn,floatfix]{revtex4}
\usepackage{graphicx}
\usepackage{color}

\begin{document}
\title{Avoided crossings between bound states of ultracold Cesium dimers}
\author{Jeremy M. Hutson}
\affiliation{Department of Chemistry, University of Durham, South
Road, Durham, DH1~3LE, United Kingdom}
\author{Eite Tiesinga and Paul S. Julienne}
\affiliation{Joint Quantum Institute, National Institute of
Standards and Technology and The University of Maryland,
Gaithersburg, Maryland 20899-8423, USA}

\date{\today}

\begin{abstract}
We present an efficient new computational method for
calculating the binding energies of the bound states of
ultracold alkali-metal dimers in the presence of magnetic
fields. The method is based on propagation of coupled
differential equations and does not use a basis set for the
interatomic distance coordinate. It is much more efficient than
the previous method based on a radial basis set and allows many
more spin channels to be included.  This is
particularly important in the vicinity of avoided crossings
between bound states. We characterize a number of different
avoided crossings in Cs$_2$ and compare our converged
calculations with experimental results. Small but significant
discrepancies are observed  in both crossing
strengths and level positions, especially for levels with $l$
symmetry (rotational angular momentum $L=8$). The discrepancies
should allow the development of improved potential models in
the future.
\end{abstract}

\maketitle

\section{Introduction}\label{Intro}

Ultracold Cs atoms are of great interest for a number of
experiments, which have produced a Bose-Einstein condensate of
such atoms~\cite{Weber2003}, formed a cold cloud of Cs$_2$
dimer molecules~\cite{Herbig2003}, probed three-body Efimov
physics~\cite{Kraemer2006}, studied collisional
shifts~\cite{Szymaniec2007} or quantum
scattering~\cite{Hart2007} of atomic clock states, carried out
high-resolution molecular spectroscopy~\cite{Vanhaecke2004} or
used magnetic fields to switch among a variety of very weakly
bound molecular states of the Cs$_2$
dimer~\cite{Mark2007b,Mark2007a}. These experiments all depend
upon and take advantage of the collisional interactions between
two Cs atoms.  Consequently, accurate theoretical and
computational models of near-threshold Cs atom scattering and
bound states are necessary for maximum understanding of
existing experiments and for making quantitative predictions
for new experimental domains.

Because of the complex spin structure of two ground-state Cs
atoms, many different near-threshold bound states exist and
have different magnetic moments. They thus tune differently
with magnetic field. When one of these bound states crosses a
collision threshold, a low-energy scattering resonance occurs,
commonly known as a Feshbach resonance.  Extensive study of
such resonances has allowed the construction of quite accurate
coupled-channel models for calculating the magnetic
field-dependent scattering and bound-state properties near
collision
thresholds~\cite{Chin2000,Leo2000,Chin2003,Chin:cs2-fesh:2004,Mark2007b,Mark2007a}.
These models incorporate the electron and nuclear spins, their
mutual interactions, and the adiabatic Born-Oppenheimer
potentials for the X$^1\Sigma_g^+$ and a$^3\Sigma_u^+$
molecular states that correlate with two $^2$S$_{1/2}$
ground-state Cs atoms.  By adjusting the model parameters to
fit the measured magnetic fields for resonances in different
scattering channels, the model quite accurately predicts
near-threshold scattering properties and the binding energies
of weakly bound states within a few GHz of threshold.  Such
threshold models can also be adapted to treat three-body
interactions, for which an accurate knowledge of the threshold
two-body bound states is necessary~\cite{Lee2007}. The models
are sensitive to relatively few parameters, and may or may not
be adequate when extended into new experimental domains.

Recently, Mark {\em et al.}~\cite{Mark2007b,Mark2007a} have
characterized a number of avoided crossings between levels
bound by only $E/h\approx 5$ MHz with respect to the energy of
two separated Cs atoms in their lowest-energy Zeeman sublevels.
Using time-dependent magnetic field ramping, they were able to
convert two Cs atoms into a number of different molecular
states with different rotational quantum numbers and magnetic
moments. Most of the bound states are well described by the
existing coupled-channel model in regions far from avoided
crossings. However, characterizing the avoided crossings
themselves presents problems for the existing computational
methods. In particular, Ref.~\cite{Chin:cs2-fesh:2004}
calculated bound states using a method based on a basis set
expansion of the radial wavefunctions in a discrete variable
representation (DVR). This method can use only a restricted
spin basis in determining the molecular bound states because of
the large number of grid points required.

The present paper develops an improved computational method
that is necessary to calculate and understand the avoided
crossings in Cs$_2$. This method uses a propagator approach
\cite{Hutson:CPC:1994} in place of a radial basis set to
represent the molecular bound states. It can readily be adapted
to threshold states of other
molecules~\cite{Kohler2006,Lang2008}. The propagator approach
is computationally much cheaper than the DVR approach and as a
result can include many more coupled spin channels. The new
approach is used to compare the calculated and observed
properties of the avoided crossings, in order to identify
aspects of the ground-state coupled-channel model for Cs$_2$
that are still in need of improvement.

\section{Computational methods}\label{CompMethods}

\begin{figure}[htbp]
\includegraphics[width=3.2in]{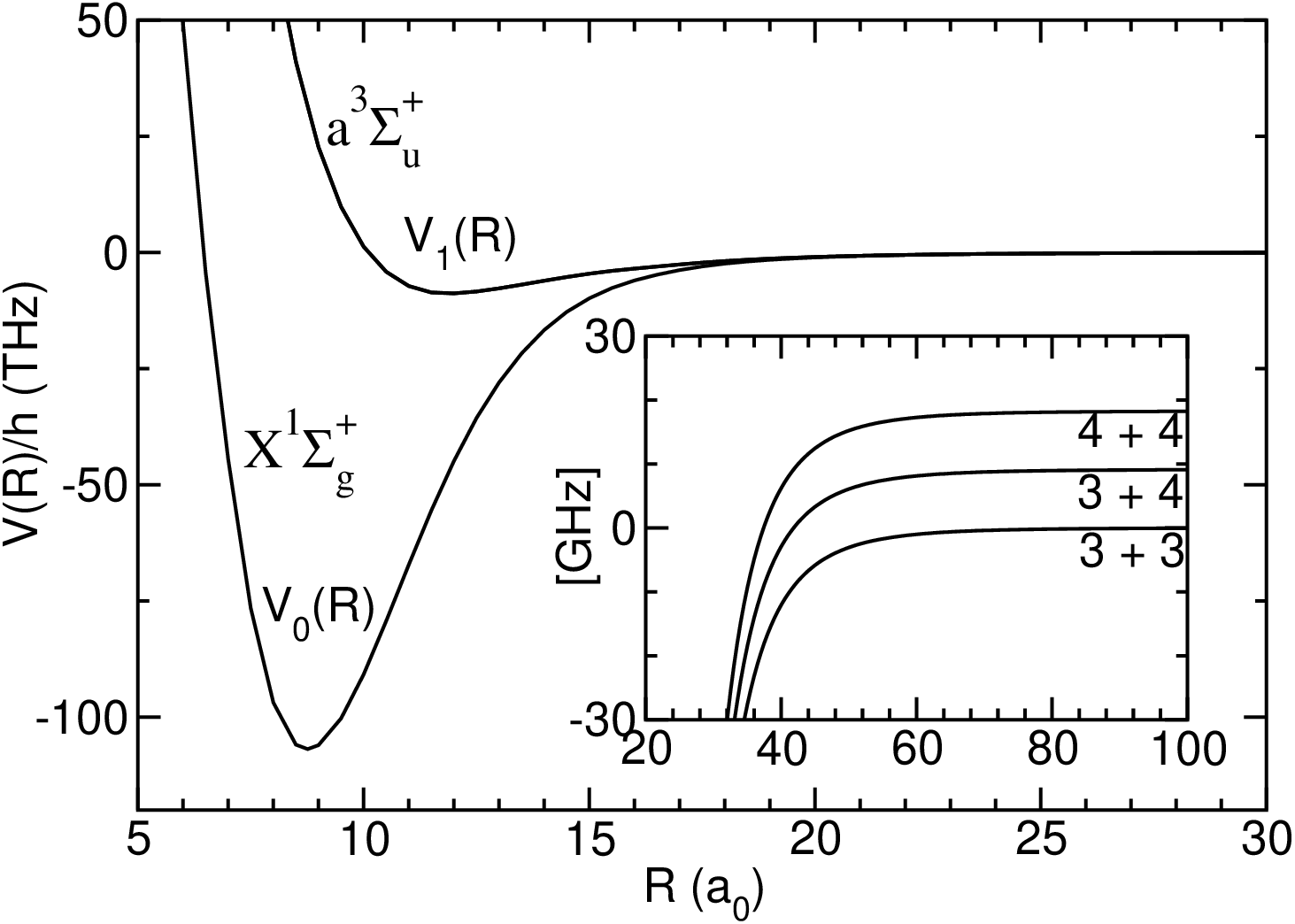}
\caption{Molecular potential energy curves $V_0(R)$ and
$V_1(R)$ for the respective singlet and triplet states of
Cs$_2$ correlating with two separated $^2$S$_{1/2}$
ground-state atoms. The inset shows an expanded view of the
long-range potentials separating to the two different $f=$ 3
and 4 hyperfine states of the $^2$S$_{1/2}$ atom with nuclear
spin $i$=$7/2$ and magnetic field $B=0$.  The inset shows the
adiabatic potentials obtained from diagonalizing the matrix
form of the operator $\hat h_1 + \hat h_2 +\hat V(R)$ at each
$R$ for the case of $L=0$, $M_F=+6$. There are 5 channels, and
the $3+4$ and $4+4$ separated-atom limits are doubly degenerate
at $B=0$.  All 5 channels have the same long-range variation as
$-C_6/R^6$, with $C_6=6860$ $E_{\rm h}
a_0^6$~\cite{Chin:cs2-fesh:2004} ($E_{\rm h}=4.3597 \times
10^{-18}$ J is the Hartree and $a_0$=0.0529177 nm is the Bohr
radius). The level crossings discussed in this paper are for
very weakly bound levels that lie within about $E/h \approx 5$
MHz of the dissociation limit to two $\{f m_f\}=\{3,+3\}$ atoms
in the magnetic field range from 0 mT to 5 mT.} \label{fig1}
\end{figure}

The present work solves the bound-state Schr\"odinger equation for
Cs$_2$ using two independent methods. In either case the
Hamiltonian may be written
\begin{equation}
\frac{\hbar^2}{2\mu} \left[-R^{-1} \frac{d}{dR^2} R + \frac{\hat
L^2}{R^2} \right] + \hat h_1 + \hat h_2 + \hat V(R),
\label{eq:SE}
\end{equation}
where $\mu$ is the reduced mass and $\hat L^2$ is the operator
for the end-over-end angular momentum of the two atoms about
one another. The monomer Hamiltonians including Zeeman terms
are
\begin{equation}
\hat h_j = \zeta \hat \imath_j \cdot \hat s_j + g_e \mu_{\rm B}
B \, \hat s_{zj} + g_n \mu_{\rm B} B \, \hat \imath_{zj},
\label{eq:h-hat}
\end{equation}
where $\hat s_1$ and $\hat s_2$ represent the electron spins of
the two atoms and $\hat \imath_1$ and $\hat \imath_2$ represent
nuclear spins. $g_e$ and $g_n$ are the electron and nuclear
$g$-factors, $\mu_{\rm B}$ is the Bohr magneton, and $\hat s_z$
and $\hat \imath_z$ represent the $z$-components of $\hat s$
and $\hat \imath$ along a space-fixed $Z$ axis whose direction
is defined by the external magnetic field $B$. The interaction
between the two atoms $\hat V(R)$ is given by Stoof {\it et
al.}~\cite{Stoof1988} as the sum of two terms,
\begin{equation}
{\hat V}(R) = \hat V^{\rm c}(R) + \hat V^{\rm d}(R)\,.
\label{eq:V-hat}
\end{equation}
Here $\hat V^{\rm c}(R)=V_0(R)\hat{\cal{P}}^{(0)} + V_1(R)\hat{
\cal{P}}^{(1)}$ is an isotropic potential operator that depends
on the potential energy curves $V_0(R)$ and $V_1(R)$ for the
respective X$^1\Sigma_g^+$ singlet and a$^3\Sigma_u^+$ triplet
states of the diatomic molecule.  The singlet and triplet
projectors $\hat{ \cal{P}}^{(0)}$  and $\hat{ \cal{P}}^{(1)}$
project onto subspaces with total electron spin quantum numbers
0 and 1 respectively.  Figure~\ref{fig1} shows the two
potential energy curves for Cs$_2$.  The $\hat V^{\rm d}(R)$
term represents small, anisotropic spin-dependent couplings
that are responsible for the avoided crossings discussed in
this paper and are discussed further in Section~\ref{CompareT}
below.

The first method for finding eigenvalues is a conventional full
matrix diagonalization in a discrete variable representation (DVR)
 \cite{Colbert:1992}.  It uses a basis set made up of products of
internal and radial functions. The internal Bose-symmetrized
basis set is made up of functions in which the operators $\hat
L^2$ and $\hat h_j$ are diagonal, that is,
\begin{equation}
|\alpha_1 m_{f1}\rangle | \alpha_2 m_{f2} \rangle |L M_L\rangle,
\label{eqbascoup}
\end{equation}
where $ |L M_L\rangle$ and $|\alpha_j m_{fj}\rangle$
respectively represent the eigenstates of $\hat L^2$ and the
$B$-dependent monomer Hamiltonian $\hat h_j$, and where $M_L$
and $m_{fj}$  are projection quantum numbers along the magnetic
field direction. When $B=0$, $|\alpha_j m_{fj}\rangle=|(s_j
i_j) f_j m_{fj}\rangle$, where $f_j$ is the total spin of atom
$j$ and $m_{fj}$ is its space-fixed projection. As $B$
increases from zero, different $f_j$ values become mixed. The
DVR radial functions are unevenly spaced
collocation points obtained from a nonlinear coordinate
transformation \cite{Tiesinga:na2:1998}.

This DVR method requires diagonalizing a large $N \times N$
matrix, the dimension of which is given by the product of the
number of spatial collocation points $N_{\rm c}$ and the number
of spin basis functions $N_{\rm s}$.  We use the LAPACK
subroutine DSPEVX to find a selected range of eigenvalues and
eigenvectors~\cite{lapack}.  In order to use a direct
diagonalization procedure to calculate the bound-state
energies~\cite{Note1} shown in
Refs.~\cite{Chin:cs2-fesh:2004,Mark2007a}, the magnitude of
$N=N_{\rm c} N_{\rm s}$ was limited to around 25000 using a
processor with 4 GB of memory.  With $N_{\rm c} \approx 800$,
in order to give 5 points per node with about 150 nodes for
threshold wave functions, the number of spin basis functions is
thus restricted to be about $N_{\rm s} \approx  35$.  When this
is fewer than is needed for a complete calculation, an
approximation scheme becomes necessary, as described in
Section~\ref{CompareT}.

The second method avoids the use of a basis set for the
interatomic distance $R$ and instead relies on propagation of
coupled differential equations \cite{Hutson:CPC:1994}. In this
case the Bose symmetrized basis set used is a fully decoupled
set,
\begin{equation} \Phi_k = |s_1 m_{s1}\rangle|i_1 m_{i1}\rangle
|s_2 m_{s2}\rangle |i_2 m_{i2}\rangle |L M_L\rangle.
\label{eqbasdecoup}
\end{equation}
The compound channel index $k$ is used to simplify notation and
implies values of all the quantum numbers in the basis set.
While the choice of the basis sets in Eqs.~(\ref{eqbascoup})
and (\ref{eqbasdecoup}) represent different approaches, they
are equivalent for representing molecular energy levels when
the two basis sets span the same space. There is a simple
unitary transformation between the two basis sets. The matrix
elements of the different terms in the Hamiltonian in basis set
(\ref{eqbasdecoup}) are given in the Appendix.

In the propagation method, we expand the total wavefunction for
state $n$ as
\begin{equation}
\Psi_n = R^{-1} \sum_k \Phi_k \psi_{kn}(R).
\end{equation}
Substituting into the Schr\"odinger equation and projecting onto
each channel function in turn gives a set of coupled equations for
the radial channel functions $\psi_{kn}(R)$,
\begin{equation}
\frac{d^2\psi_{jn}}{dR^2} = \sum_k \left[W_{jk}(R) - \varepsilon
\delta_{jk} \right] \psi_{kn}(R), \label{eqcoup}
\end{equation}
where $\delta_{jk}$ is the Kronecker delta, $\varepsilon$ is the
energy $E$ scaled by $2\mu/\hbar^2$, and
\begin{equation}
W_{jk}(R) = \int \Phi_j^* \left[ \frac{\hat L^2}{R^2} +
\frac{2\mu}{\hbar^2} \left( \hat h_1 + \hat h_2 + \hat V(R)
\right) \right] \Phi_k\,d\tau,
\end{equation}
where $d\tau$ indicates integration over all coordinates except
$R$. If there are $N_{\rm s}$ basis functions, the required
solution $\psi_n(R)$ is a column vector of order $N_{\rm s}$
with elements $\psi_{kn}(R)$. However, Eq.\ \ref{eqcoup} has
$N_{\rm s}$ independent solution vectors at any energy, so that
until the boundary conditions are applied $\psi_n(R)$ is an
$N_{\rm s}\times N_{\rm s}$ wavefunction {\em matrix}.

The Schr\"odinger equation can be solved to find an $N_{\rm
s}\times N_{\rm s}$ wavefunction matrix at any energy $E$. In
practice it is numerically stabler to propagate the
log-derivative matrix $Y(R)=[d\psi_n/dR][\psi_n(R)]^{-1}$.
However, a solution that satisfies bound-state boundary
conditions can be found only at the eigenvalues $E_n$.
Solutions are propagated outwards from a point $R_{\rm min}$ in
the inner classically forbidden region and inwards from a point
$R_{\rm max}$ at long range to a matching point $R_{\rm mid}$.
The outwards and inwards solutions are designated $Y^+(R)$ and
$Y^-(R)$. If $E$ is an eigenvalue of the coupled equations,
there must exist a wavefunction vector $\psi_n(R_{\rm mid}) =
\psi_n^+(R_{\rm mid}) = \psi_n^-(R_{\rm mid}$) for which the
derivatives also match,
\begin{equation}
\left.\frac{d\psi_n^+}{dR}\right|_{R_{\rm mid}} =
\left.\frac{d\psi_n^-}{dR}\right|_{R_{\rm mid}},
\end{equation}
so that
\begin{equation}
Y^+(R_{\rm mid}) \psi_n(R_{\rm mid}) = Y^-(R_{\rm mid})
\psi_n(R_{\rm mid}).
\end{equation}
Thus $\psi_n(R_{\rm mid})$ is an eigenvector of $Y^+(R_{\rm mid})
- Y^-(R_{\rm mid})$ with eigenvalue 0. It is thus possible to
locate eigenvalues of the Schr\"odinger equation by propagating
solutions of the coupled equations and searching for zeroes in the
eigenvalues of the log-derivative matching matrix $Y^+(R_{\rm
mid}) - Y^-(R_{\rm mid})$ as a function of energy. This approach
is much stabler for large multichannel problems than the older
approach \cite{Johnson:1978} of searching for zeroes of the {\em
determinant} of the matching matrix.

The major advantage of the propagator method is that the
matrices handled are only of dimension $N_{\rm s}\times N_{\rm
s}$, where $N_{\rm s}$ is the number of {\em internal} basis
functions. The computational cost is proportional to $N_{\rm
s}^3$ but only {\em linear} in the number of propagation steps.
By contrast, a full diagonalization with $N_{\rm c}$ radial
basis functions (collocation points) involves matrices of
dimension $N_{\rm s}N_{\rm c} \times N_{\rm s}N_{\rm c}$. The
computational cost is proportional to $N_{\rm s}^3 N_{\rm
c}^3$. Since $N_{\rm c}$ typically needs to be greater than 500
for the present application, the propagator approach is much
cheaper.

The BOUND program \cite{Hutson:bound:1993} is a general-purpose
package to solve the bound-state Schr\"odinger equation using
propagator methods. The algorithms used are described in more
detail in Ref.\ \onlinecite{Hutson:CPC:1994}. For the purpose
of the present work we have generalised the BOUND package in
three significant respects:
\begin{enumerate}
\item We have generalised the structure of the code so that it can
handle coupled equations in basis sets that are not diagonal at
$R=\infty$; \item We have implemented the specific set of coupled
equations required for Cs$_2$ with the basis set of Eq.\
\ref{eqbasdecoup}; \item We have added an option to use the
log-derivative propagator of Alexander and Manolopoulos
\cite{Alexander:1987}, which is based on Airy functions and allows
very large step sizes at long range.
\end{enumerate}

In the presence of a magnetic field, the only rigorously
conserved quantum numbers are $M_{\rm tot} = m_{f1}+m_{f2}+M_L
= m_{s1}+m_{s2}+m_{i1}+m_{i2}+M_L$ and the total parity
$(-1)^L$. This leads to an infinite number of channels.
However, $L$ and $M_F=m_{f1}+m_{f2}$ are very good approximate
quantum numbers because the only term in the Hamiltonian that
is off-diagonal in them is the small anisotropic coupling term
$\hat V^{\rm d}$. In either computational approach it is
possible to restrict the number of channels by selecting only
one or a few values of $L$ and all or a subset of possible
$M_F$ values. Here we consider the case studied experimentally
by Mark {\it et al.}~\cite{Mark2007a}, who used Cs atoms is
their lowest energy hyperfine state with $m_f=+3$ to make
Cs$_2$ molecules with $M_{\rm tot}=+6$. The number of channels
with $L=0$, 2, 4, 6, 8, including all allowed $M_F$ values, are
5, 23, 46, 76, 103, respectively. Thus, for example, a full
calculation including all channels with $L=4$, 6 and 8 requires
225 channels.

In practical terms, for example, a run with the DVR method to
find 28 bound states within 3 GHz of the $E=0$ threshold for
Cs$_2$ for a single magnetic field with 30 channels and 720
collocation points took about 7 hours on an 2.4 GHz processor.
With the propagator approach we were able to find selected
near-dissociation levels for 30 channels in about 40 seconds
per level with a 2.0 GHz processor.   The great advantage of
the propagator approach was demonstrated by our ability to find
levels with 225 channels in about 45 minutes per level.  A
calculation with 225 channels would not be possible at all
using the DVR method with a direct eigenvalue solver.

\section{Comparison of computational results}\label{CompareT}

The DVR and propagator calculations described here both use the
same potential model, with the parameters given by Chin {\it et
al.}~\cite{Chin:cs2-fesh:2004}.  The potential energy curves
are based on the {\em ab initio} calculations of Krauss and
Stevens \cite{Krauss:1990}.  The singlet and triplet scattering
lengths $a_{\rm S}$ and $a_{\rm T}$, the long-range
coefficients $C_6$ and $C_8$, and a scaling factor $S_C$ for
the second-order spin-orbit coupling were adjusted by Chin {\it
et al.}\ to reproduce a substantial number of Feshbach
resonances with $L\le 4$.

\begin{figure}[htbp]
\includegraphics[width=3.2in]{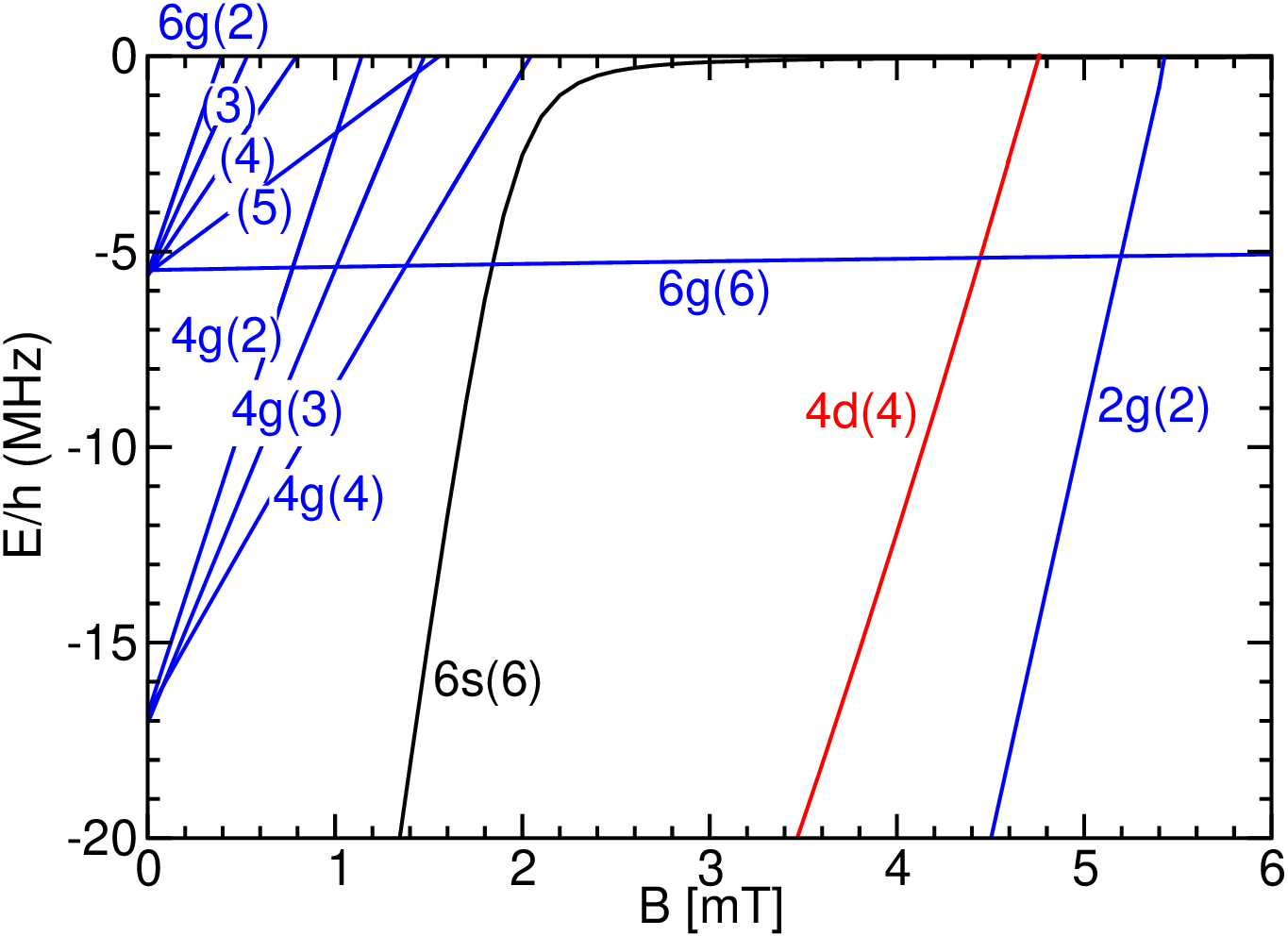}
\caption{Bound state energy $E/h$ as a function of $B$ for
levels of the Cs$_2$ molecules with even $L \le 4$ and
$M_\mathrm{tot}=+6$.  Energies are given relative to the energy
of two Cs atoms in their ground Zeeman sublevel ($f=3$,
$m_f=+3$).  The $FL(M_F)$ labeling scheme is shown for each
level.  Off-diagonal coupling between levels with different
$FL(M_L)$ quantum numbers is neglected in this calculation.}
\label{fig2}
\end{figure}

Figure~\ref{fig2} shows an example of weakly bound levels of
the Cs$_2$ molecule with $M_\mathrm{tot}=+6$ in the 0 mT to 6
mT range of $B$. Many of these levels have been probed in the
experiment of Mark {\it et al.}~\cite{Mark2007a}. The figure
also shows the bound-state classification scheme of Chin {\it
et al.}~\cite{Chin:cs2-fesh:2004}, namely $FL(M_F)$, where $F$
is the resultant of the separated-atom spins $f_1$ and $f_2$
and $M_F$ is its projection defined above.  Like $f_1$ and
$f_2$, $F$ is a good  approximate quantum number for labeling
near-threshold levels at low $B$. Quantum numbers $L=0$, 2, 4,
6, 8 are represented by labels $s$, $d$, $g$, $i$, $l$,
respectively. $M_L$ need not be specified since $M_L =
M_\mathrm{tot}-M_F$.   Fig.~\ref{fig2} shows levels with $L \le 4$
obtained from a DVR calculation that included only basis
functions for a single $L$ and $M_F$.  This neglects the small
off-diagonal couplings between levels with different $L$ and
$M_F$ quantum numbers due to $\hat V^{\rm d}$, so that levels
of different symmetry show crossings rather than avoided
crossings in Fig.~\ref{fig2}.

For ground-state alkali-metal atom interactions, the $\hat
V^{\rm d}$ operator has the form of spin-dipolar coupling
\begin{equation}
\label{eq:Vd} \hat V^{\rm d}(R) = \lambda(R) \left ( \hat s_1\cdot
\hat s_2 -3 (\hat s_1 \cdot \vec e_R)(\hat s_2 \cdot \vec e_R)
\right ) \,,
\end{equation}
where $\vec e_R$ is a unit vector along the internuclear axis and
$\lambda$ is an $R$-dependent coupling constant, which for our
model is
\begin{equation}
\label{eq:lambda}
\lambda(R) = E_{\rm h} \alpha^2 \left ( \frac{1}{(R/a_0)^3}
- 0.071968 e^{-0.83[(R/a_0)-10]} \right ) \,,
\end{equation}
where $\alpha\approx 1/137$ is the fine structure constant. At
large $R$ the coupling becomes the long-range dipolar
interaction between the spins on the separated atoms that
varies as $1/R^3$~\cite{Stoof1988, Moerdijk1995}.  In the
short-range region of chemical bonding the magnitude of
$\lambda(R)$ is primarily determined by the second-order
spin-orbit coupling term represented by the exponential
term~\cite{Mies1996,Leo:1998,Kotochigova2001,Chin:cs2-fesh:2004}.

The crossings in Fig.~\ref{fig2} become avoided crossings when
the small interactions due to $\hat V^{\rm d}$ are taken into
account. The energy splitting at the crossing varies greatly,
depending on the quantum numbers of the two levels.  In first
order, the $\hat V^{\rm d}$ operator couples states $FL(M_F)$
and $F'L'(M_F')$ according to the selection rules $|L-L'|=$ 0
or 2, $|F-F'|=$ 0, 1, or 2, and $|M_F-M_F'|=$ 0, 1, or 2. These
selection rules immediately follow from the tensor form of the
operator in Eq.~(\ref{eq:Vd}), as given by Stoof {\it et
al.}~\cite{Stoof1988}, who write Eq.~(\ref{eq:Vd}) as a sum of
products of $L=2$ spherical harmonic components $Y_{LM_L}({\vec
e_R})$ and rank 2 spin tensor components.  We refer to a crossing as 
direct when there is a first order coupling of the two states
involved and indirect when there is not.

The success of a calculation of the Cs$_2$ energy levels and
their avoided crossings depends on the sufficiency of the basis
set expansion of the wave function.  Suppose we wish to
calculate the energy of one $FL(M_F)$ state that crosses a
different $F'L'(M_F')$ state.  It is necessary to include
sufficient basis functions to represent each state adequately,
and to represent their interaction.  This is simplified by
taking advantage of the selection rules described above.  In
order to represent a level with a given $FL(M_F)$, it is
necessary to include all basis functions with the same set of
three quantum numbers, since such levels are coupled by terms
due to the strong central potential $\hat V^{\rm c}$.   A level
calculated with such a basis is coupled through the $\hat
V^{\rm d}$ operator to other levels in which one or more of the
three quantum numbers are different.  Such off-diagonal
coupling causes shifts in level positions and also induces
avoided crossings.

In the propagator calculations, the basis set usually includes
all functions with $L$ and $L'$ of the levels in question
consistent with $M_\mathrm{tot}$. Additional basis functions
with different quantum numbers $L_i$ are added to account for
shifts and crossings due to coupling of $L$ or $L'$ with $L_i$.
The propagator basis is specified by giving $L$, $L'$, and a
list of additional values $L_i$ needed to account for
higher-order coupling.  In the DVR calculations, the basis sets
are additionally limited by restricting the calculation to
functions with $L(M_F)$, $L'(M_F')$ and additional quantum
numbers $L_i(M_{F,i})$ as needed.  Thus the basis set is
specified by giving the list $L(M_F)L'(M_F')[L_i(M_{F,i})]$.
Some propagator calculations were done with a similarly
restricted list to verify that the two methods gave exactly
equivalent results. Neither the propagator nor DVR calculations
make any additional restrictions by $F$, although this could be
done.

\begin{figure}[htbp]

\includegraphics[width=3.2in]{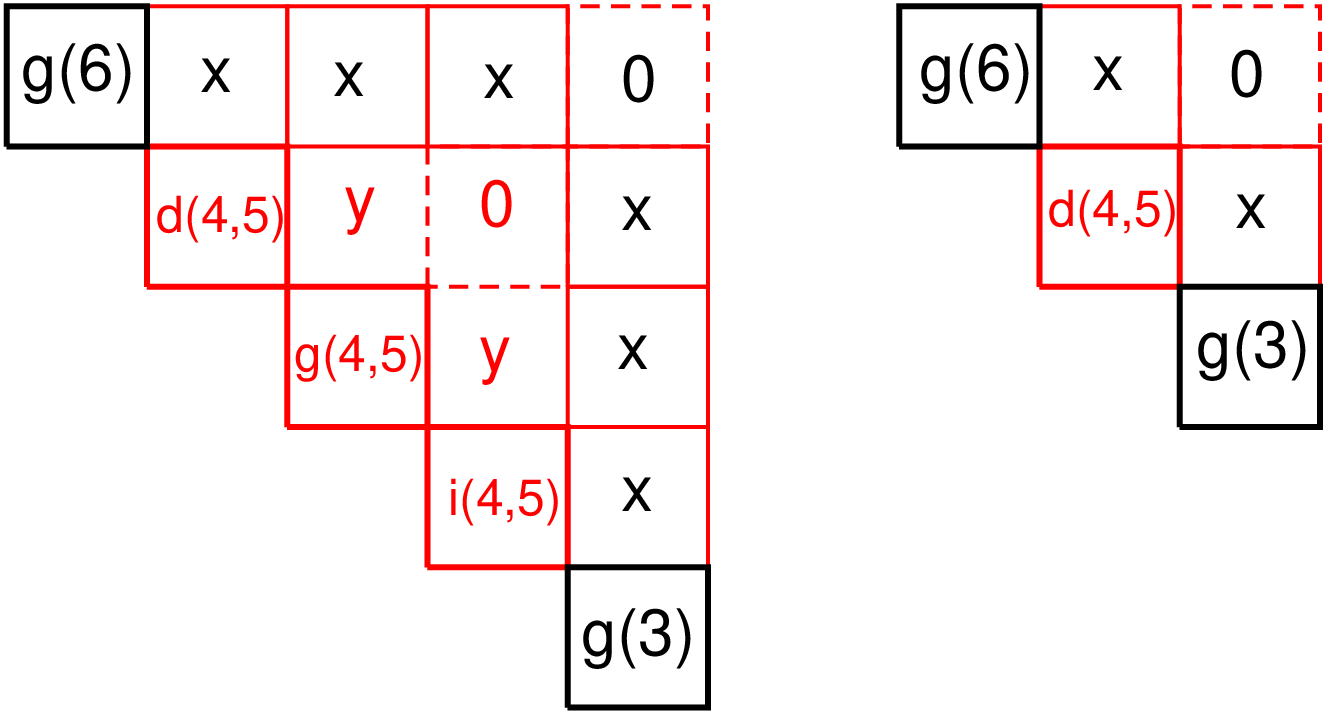}
\caption{Example of coupling between different $L(M_F)$
symmetry blocks with the symmetry of the dipole-dipole
interaction of $\hat V^{\rm d}$. Each block represents a
Hamiltonian matrix for spin states with the $L(M_F)$ values
indicated.  The labels ``x'' and ``y'' indicate the existence
of nonvanishing coupling due to $\hat V^{\rm d}$; a ``0''
indicates no coupling. The case shown is for a $g(6)$ and a
$g(3)$ level, which have no direct coupling. The left panel
shows the symmetries that give rise to second-order
interactions between the two levels and thus contribute to the
strength of the avoided crossing between them.  The right panel
shows a truncated set of interactions through intermediate
$d(4)$ and $d(5)$ levels.} \label{fig3}
\end{figure}

\begin{figure}[htbp]
\includegraphics[width=3.2in]{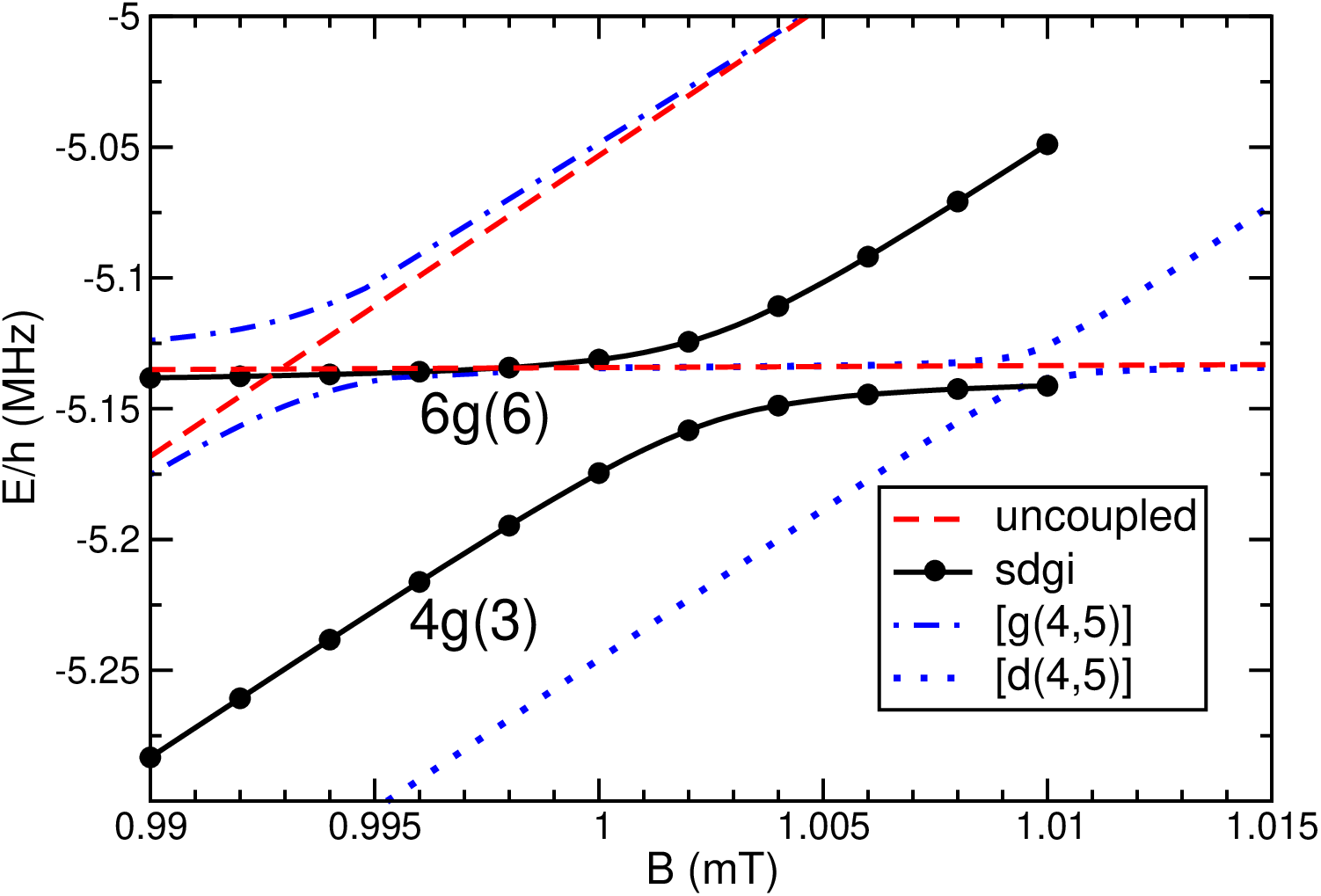}
\caption{Calculated energy levels $E/h$ with
$M_\mathrm{tot}=+6$ as a function of magnetic field $B$ near
the crossing of the $4g(3)$ level with the 6g(6) level near 1.0
mT.  The points and solid line show the propagator calculation with
a $sdgi$ basis set.  The dashed lines show the crossing levels
from two uncoupled DVR calculations with $g(3)$ or $g(6)$ basis
functions only. The dash-dot and dotted lines show the crossing
levels from DVR calculations with added $g(4,5)$ and $d(4,5)$
functions respectively.  The DVR calculation with $i(4,5)$
basis functions is not shown, but lies near the uncoupled
crossing and has a very small splitting, indicating very weak
second-order coupling through distant $i$ states.} \label{fig4}
\end{figure}

Figure~\ref{fig3} illustrates the size of the basis set needed,
as governed by the selection rules on $\hat V^{\rm d}$
coupling.   Since the matrix elements of $\hat V^{\rm d}$ are
relatively small, they are normally of practical significance
only through second order.  Thus it is necessary to include
only intermediate levels with $L_i$ and $M_{F,i}$ that differ
from $L$ or $L'$ and  $M_F$ or $M_F'$ by at most 2 units.  Any
higher-order couplings would be much smaller than those
discussed here. Thus, in order to represent the crossing of a
$6g(6)$ and a $4g(3)$ level, for which there is no first-order
direct coupling, $d$-, $g$-, or $i$-basis functions with $M_F=$
4 and 5 need to be included in the basis, as shown in
Fig.~\ref{fig3}.  To represent additional second-order shifts
of the two $g$ levels, $d$-, $g$- and $i$-basis functions with
$M_F=$ 1, 2, 7 and 8 also need to be added.

Figure~\ref{fig4} illustrates calculations with different basis
sets, comparing energies calculated with the propagator and DVR
methods for the crossing of the $4g(3)$ and $6g(6)$ levels near
1.0 mT.  Table~\ref{tab:crossings} tabulates the positions and
strengths of this crossing, as well as a number of others. The
position $B_0$ of the crossing is defined as the field at which
the two levels are closest together and the strength $2V$ is
the minimum of the difference between the two energies as a
function of $B$; $2V$ is used since the splitting is twice the
effective coupling matrix element $V$ in a 2-level
representation of the crossing~\cite{Mark2007a}. We have
verified that the two methods give identical results within
numerical accuracy when exactly equivalent basis sets are used.
Since there is no direct interaction between the two crossing
levels in this case, the splitting at the crossing originates
principally in second-order interactions mediated through
distant levels of $d$, $g$, or $i$ symmetry with $M_F=$ 4 or 5.
However, as mentioned above, second-order couplings to levels
with other $M_F$ values can cause additional shifts. Both bound
and scattering states can contribute, and the contribution from
any given distant state varies inversely with the its
separation in energy from the crossing. Intermediate $g$ levels
are the closest in energy to the crossing, whereas intermediate
$i$ levels are the most distant. In Figure~\ref{fig4}, the
$sdgi$ basis set used in the propagator calculation is
effectively complete. It may be seen that a calculation
including only the $g(4,5)$ intermediate states captures most
of the crossing strength but does not reproduce the level
shifts well. Conversely, a calculation including only the
$d(4,5)$ states gives a crossing strength that is much too
small but overestimates the level shifts. The contributions to
the crossing strength from different intermediate states are
far from additive. There are no experimental results for this
crossing.

\begin{figure}[htbp]
\includegraphics[width=3.2in]{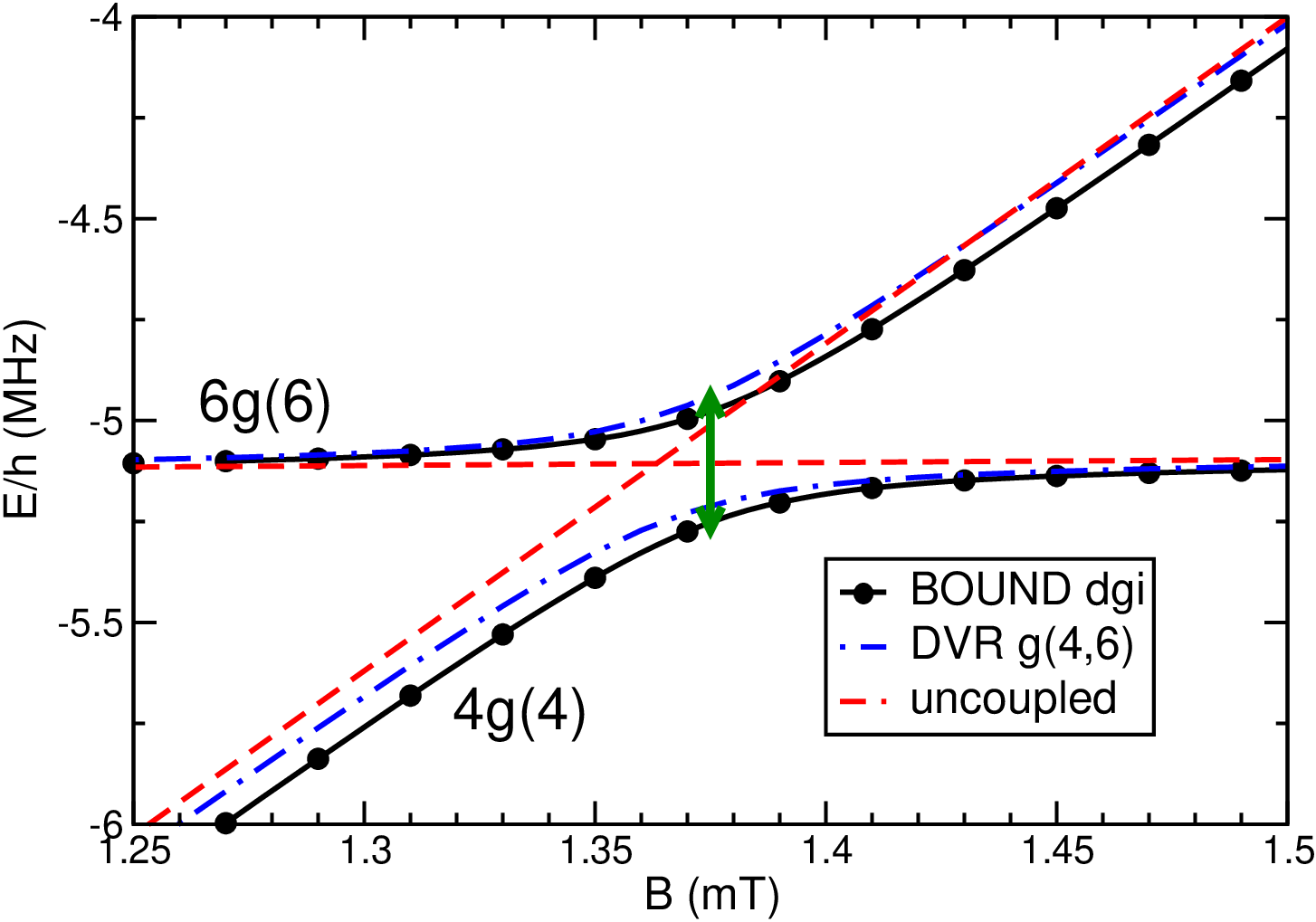}
\caption{Calculated energy levels $E/h$ with $M_\mathrm{tot}=+6$
as a function of magnetic field $B$ near the crossing of the $4g(4)$
level with the 6g(6) level near 1.0 mT. The points and solid
line show the propagator calculation with a $dgi$ basis set.  The
dashed lines shows the crossing levels from two uncoupled
DVR calculations with $g(4)$ or $g(6)$ basis functions only.
The dash-dot lines show the avoided crossing from a DVR
calculation with only the direct coupling in the $g(4,6)$ basis
set included. The doubled-headed arrow at the position of the
propagator crossing shows the measured splitting~\cite{Mark2007a}.
The actual experimental crossing was observed $0.046$ mT
lower in $B$ value than the propagator crossing.} \label{fig5}
\end{figure}

\begin{figure}[htbp]
\includegraphics[width=3.2in]{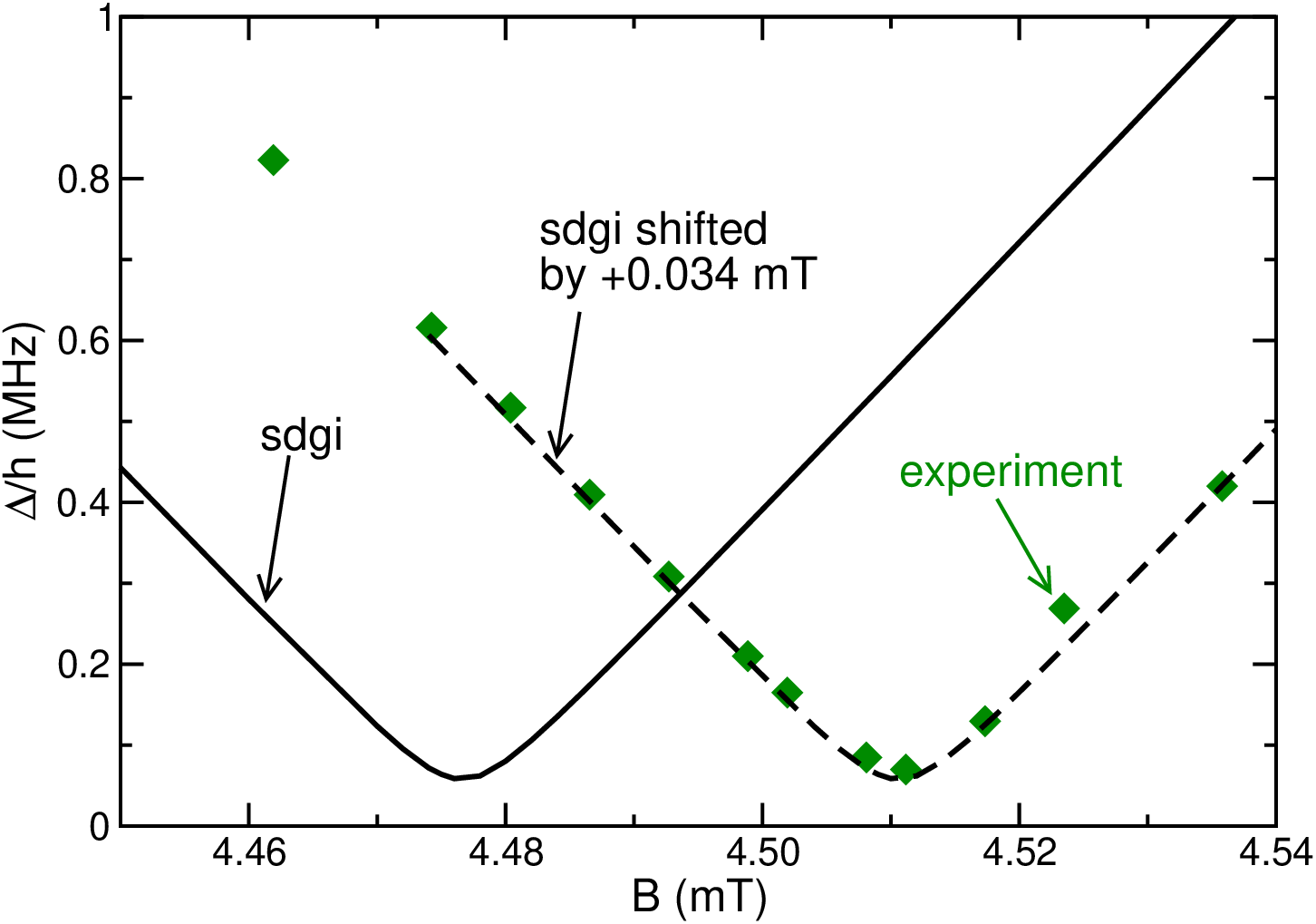}
\caption{Calculated energy difference $\Delta/h$ between the
$6g(6)$ and $4d(4)$ levels with $M_\mathrm{tot}=+6$ as a
function of magnetic field $B$.  The solid line is from a propagator
calculation with the $sdgi$ basis. The diamonds show the
experimental results obtained by Ferlaino {\it et al.} using
their more accurate field modulation
method~\cite{Ferlaino:wiggle:2008}. The dashed line shows the
calculated points shifted by $+0.034$ mT. The DVR calculation
(not shown) with direct coupling included in the $d(4)g(6)$
basis is virtually identical to the dashed line when the DVR
results are shifted by $+.062$ mT.} \label{fig5b}
\end{figure}

Figure~\ref{fig5} illustrates a different case, a $4g(4)-6g(6)$
crossing with a splitting that is about 8 times larger at the
crossing. This is a case where the two states involved have a
direct coupling to one another through $\hat V^{\rm d}$. While
additional second-order coupling can change the position and
strength of the crossing slightly, the direct coupling is
dominant and the restricted-basis DVR calculation agrees much
better with the full propagator calculations.  Both are in
reasonable agreement with the measured splitting of the
crossing~\cite{Mark2007a}. However, the calculated position in
$B$ needs to be shifted by $-0.046$ mT to agree with the
measured position~\cite{Mark2007a}.  This remaining discrepancy
reflects a real deficiency in the parameters of our potential
model as discussed below.

Figure~\ref{fig5b} shows the difference between the upper and
lower branches of the crossing for the case of the
$4d(4)-6g(6)$ crossing near 4.5 mT.  This is another case of
direct coupling, where the measured and calculated crossings
agree well in coupling strength, although the calculated
position needs to be shifted by $+0.034$ mT to agree with the
measured one.

Table~\ref{tab:crossings} show comparisons between the
propagator and DVR calculations for a number of other crossings
in Figure~\ref{fig2}.  Crossing positions generally agree
within about 0.01 mT among the different basis sets. Relatively
good agreement between propagator and DVR coupling strengths
$2V$ is seen in the cases where there is direct coupling
between the two crossing levels, or where the DVR method
includes all second-order intermediate states allowed by the
symmetry of the $\hat V^{\rm d}$ operator.  However, for
higher-$L$ crossings it was usually necessary to select a
subset of the allowed intermediate states to make DVR
calculations feasible. In such cases the DVR method can give
unreliable results, depending on the choice of restricted basis
set. 

\begin{figure}[htbp]
\includegraphics[width=3.2in]{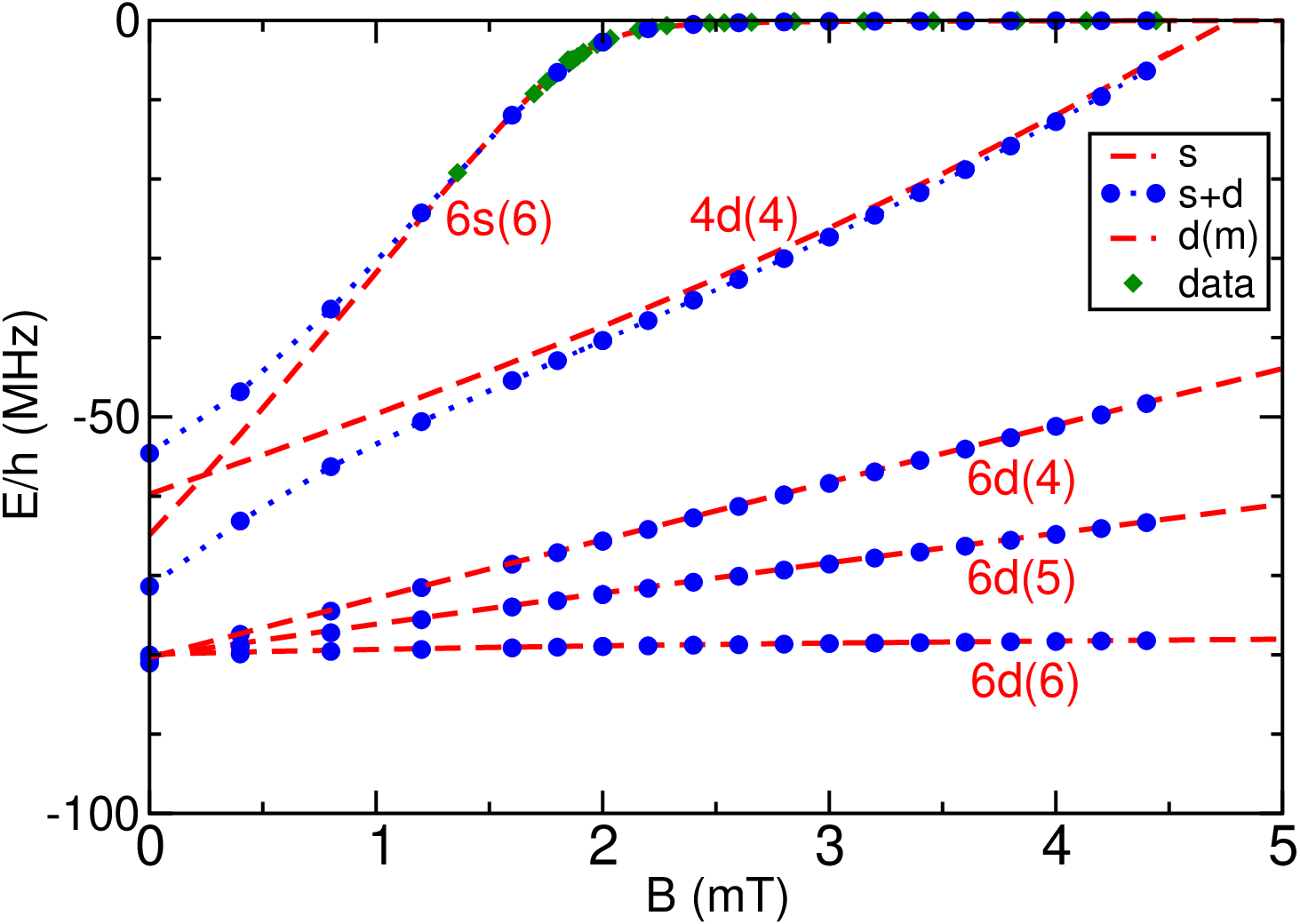}
\caption{Energy levels $E/h$ as a function of $B$ for the
Cs$_2$ molecule for $L=0$ and 2 only with $M_\mathrm{tot}=+6$
($g$ and $l$ levels are not shown).  The dashed lines show the
DVR levels calculated with the uncoupled $s(6)$ and $d(M_F)$
basis sets, where $M_F=4$, 5, or 6.  The diamonds show the results
of Mark {\em et al.}~\cite{Mark2007a}. The $4d(4)$ level crosses the
$6s(6)$ level twice, near 0.24 mT and 4.77 mT.  The closed
circles and dotted lines show the levels obtained
from a DVR calculation with an $sd$ basis.   A
propagator calculation with a full $sdg$ basis shows negligible
differences for this case.} \label{fig6}
\end{figure}

Figure~\ref{fig6} shows calculated bound states for $s$ and $d$
levels on a broader energy scale. (Levels with other symmetries
are not shown). A DVR calculation with a full $sd$ basis is
possible in this case.  The $6s(6)$ and $4d(4)$ uncoupled
levels show two crossings.   
The low-field crossing around 0.24 mT occurs near the
observed location (0.72 mT) of a three-body Borromean state
of the Cs$_3$ trimer associated with the exotic Efimov physics
of this species~\cite{Kraemer2006}. Lee {\it et
al.}~\cite{Lee2007} used the last two $6s(6)$ two-body states
of the Cs$_2$ dimer to construct the parameters for full
three-body calculations of bound states and recombination
coefficients in the 0 mT to 3 mT range.  While their method was
able to give semi-quantitative agreement with the measurements,
the avoided crossing of the $6s(6)$ level with the $4d(4)$
level needs to be taken into account in subsequent calculations
because of the mixed spin character of the target molecular
state produced by the three-body recombination in this region
of $B$.   The strong $s-d$ interactions modify the $s$-wave
scattering length at small $B$, but this is easy to take into
account by including $s$ and $d$ basis functions in scattering
calculations.

The higher-field $6s(6)-4d(4)$ crossing near 4.8 mT has been
studied in Refs.~\cite{Mark2007a,Julienne2004}.
Figure~\ref{fig7} shows an expanded view of the
very-near-threshold region of this crossing and the additional
$6s(6)-2g(2)$ crossing near 5.4 mT.   The interaction between
$s$ and $d$ states results in an overall shift in the binding
energy of the $6s(6)$ level, where the uncoupled level is too
high in energy. This case illustrates one advantage of the
propagator method over the DVR method.  The latter has to use a
finite range of spatial points and is restricted by the length
of the ``box'' in which the calculation is carried out.  When
this length is too large, the number of the spatial collocation
points can become too  large for practical calculations.  A
5000 $a_0$ ``box'' is sufficient for levels with binding
energies on the order of 40 kHz, since the scattering length,
which gives an indication of the ``size'' of the weakly bound
molecular levels~\cite{Kohler2006}, is on the order of 1000
$a_0 \ll$ 5000 $a_0$. Such restrictions on spatial grid do not
apply to the propagator method, which is capable of calculating
levels arbitrarily close to $E=0$, as long as the propagation
is to sufficiently large distances. Since the propagator used
can take very large steps at long range, this presents no
difficulty.

\begin{figure}[htbp]
\includegraphics[width=3.2in]{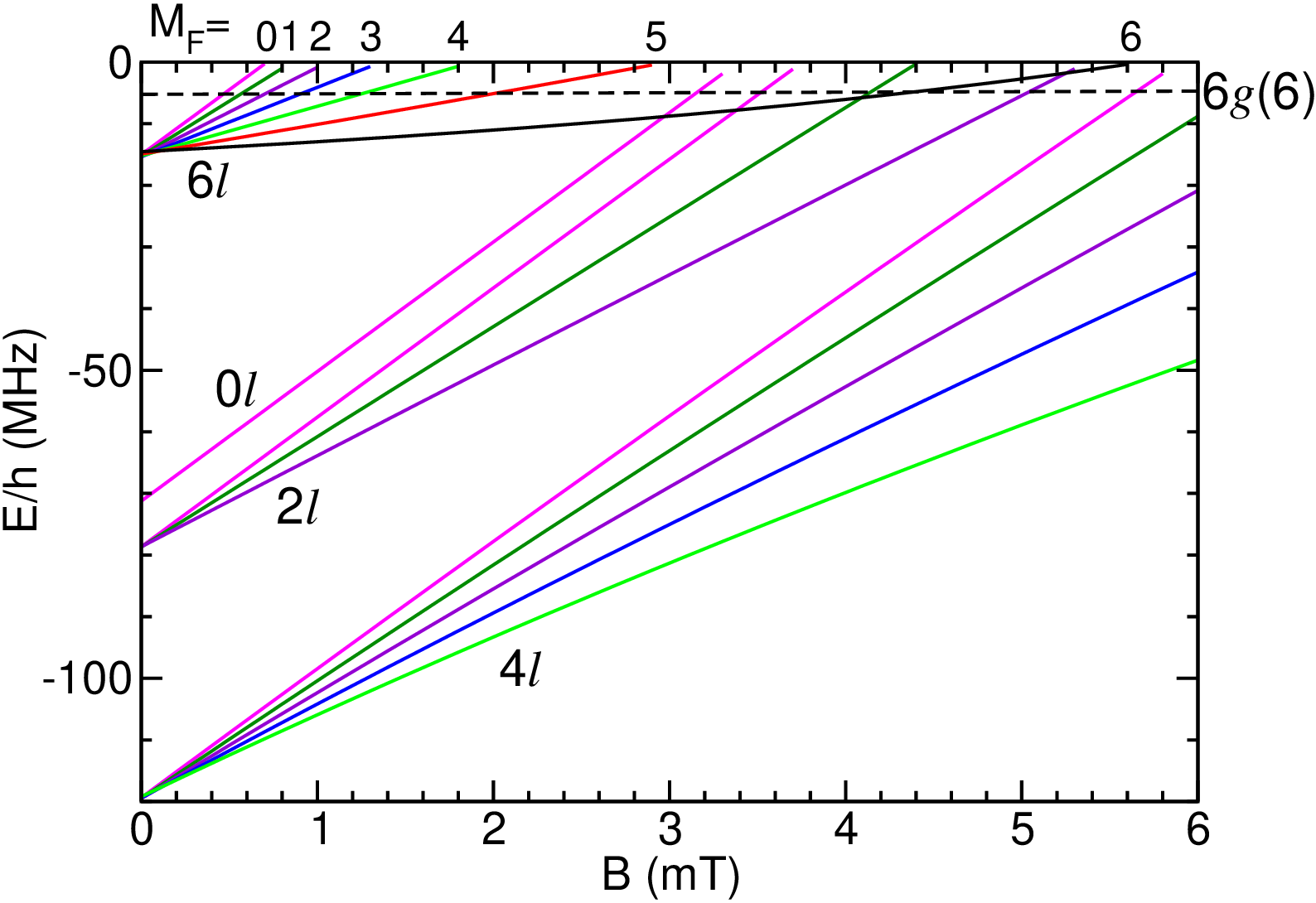}
\caption{Energy levels $E/h$ as a function of $B$ for the
Cs$_2$ molecule for $L=8$ only with $M_\mathrm{tot}=+6$.  The
solid lines show the DVR levels calculated with the uncoupled
$l(M_F)$ basis sets, where $M_F=0,\ldots,6$. The dashed line
shows the $6g(6)$ level for which avoided crossings have been
calculated (see Table~\ref{tab:crossings}) and
measured~\cite{Mark2007a}. } \label{fig6b}
\end{figure}

\begin{figure}[htbp]
\includegraphics[width=3.2in]{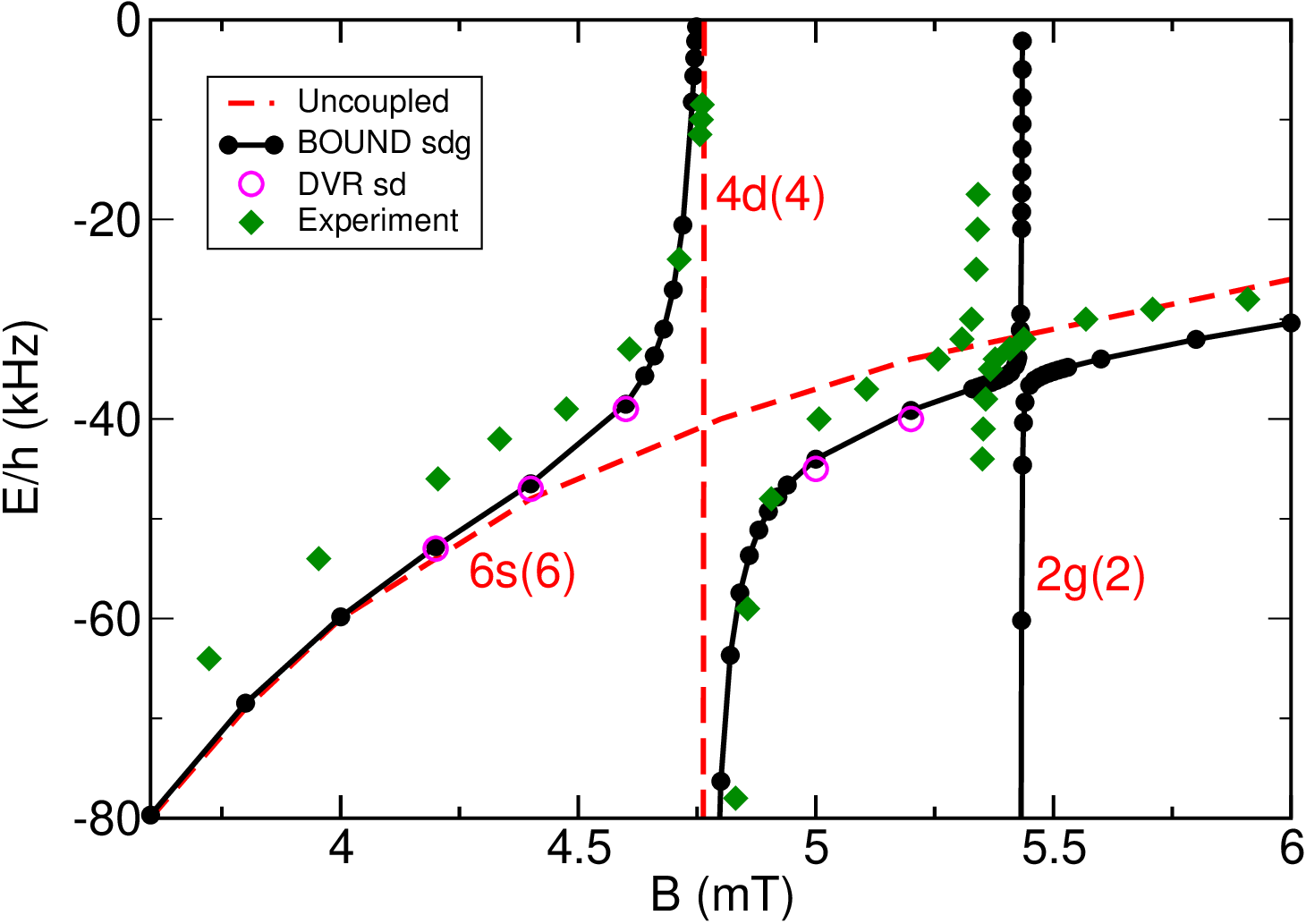}
\caption{Expanded view of the crossing in Fig.~\ref{fig6} of
the $4d(4)$ and $6s(6)$ levels near 4.8 mT.  The long dashed
line shows the uncoupled calculation with the $s(6)$ and $d(4)$
basis sets.  The solid lines show the propagator calculations with an
$sdg$ basis.  The upper crossing near 5.4 mT is due to a $2g(2)$
level.  The open circles show DVR calculations with a full
$sd$ basis in a finite box of $5000$ $a_0$.  The diamonds show
experimental results of
Lange {\it et al.}~\cite{Lange:2008}.}
\label{fig7}
\end{figure}

\begin{table}[tbhp]
\caption{Levels crossing the $6g(6)$ level of the Cs$_2$
molecule with a binding energy near $-5$ MHz relative to the
energy of two atoms in their lowest energy hyperfine state at
each $B$.  The columns label the symmetry of the crossing
state, the computational method (propagator or DVR),  the $L$
functions in the basis set used for the calculation (only
intermediate basis states are listed for the DVR method, since
basis states for the two crossing states are automatically
included), and the position $B_0$, energy $E/h$, and splitting
$2V/h$ of each crossing.  } \label{tab:crossings}
\begin{tabular}{crcrrr}
\hline\hline
State & Method & Basis  & $B_0$ & $E/h$ & $2V/h$ \\
       &                 &            &   [mT]    & ~~[MHz] & ~[kHz]  \\
\hline

4g(2)  & propagator & $g$        & 0.7615   &  -5.180   & 3.6          \\
       & propagator & $dg$       &  0.7670  &  -5.192   & 1.0          \\
       & propagator & $sdgi$     &  0.7617  &  -5.152   & 1.5          \\
       & DVR & $g(2,6)[d(4)]$ & 0.7745 &  -5.152   & 2.3          \\
       \hline
6l(3)  & propagator & $gil$      &  0.9181  &  -5.135   & 21.2         \\
       & DVR & $g(6)l(3)[i(4,5)]$  & 0.9112  &  -5.138  & 21.7    \\
       \hline
4g(3)  & propagator & $sdg$      & 1.0105  & -5.169 &  37.3            \\
       & propagator & $sdgi$     & 1.0024  & -5.138 &  33.5            \\
       &  DVR & $g(3,6)[d(4,5)]$ &  1.0097  &  -5.134  &  12.1    \\
       &  DVR & $g(3,6)[g(4,5)]$ &  0.9935  &  -5.134  &  32.9    \\
       &  DVR & $g(3,6)[i(4,5)]$ &  0.9937  &  -5.1314  &  1.1    \\
       \hline
6l(4)  & propagator & $gil$     &  1.2715 & -5.088    & 43.6     \\
\hline
4g(4) & propagator & $g$        &  1.368  & -5.14     &  264.    \\
       & propagator & $dgi$     &  1.375  & -5.11     &  277.    \\
       & DVR & $g(4,6)$    &  1.367  & -5.10     &  265.    \\
       \hline
6s(6)  & propagator & $sdg$     &  1.8648 & -5.114    & 44.8     \\
   & propagator &  $sdgi$       &  1.8664 & -5.079    & 44.8     \\
      \hline
6l(5)  & propagator & $gil$     &  2.0089 & -5.056    & 66.6     \\
\hline
6l(6)  & propagator & $gil$     &  4.3211 & -4.890    & 77.6     \\
\hline
4d(4) & propagator & $dg$       &  4.4403 & -4.937    & 55.5     \\
       & propagator & $sdgi$    &  4.4766 & -4.885    & 57.6     \\
       &  DVR & $d(4)g(6)$ &  4.4485 & -4.905    & 53.0     \\
       \hline
2g(2) & propagator & $g$        &  5.1906 & -4.887    & $<$ 0.1  \\
       & propagator & $dgi$     &  5.1176 & -4.842    & 9.5      \\
\hline\hline
\end{tabular}\\
\end{table}

\section{Comparison with experiment}\label{CompareE}

When the basis set is sufficiently large, there is good overall
agreement between our calculations and the experimental
measurements, as already noted in relation to Figs.~\ref{fig5}
and~\ref{fig5b}. Table~\ref{tab:experiment} lists other
examples, including the crossings of the $6g(6)$ level with the
$6l(M_F)$ levels shown in Fig.~\ref{fig6b}.  Since the
potentials and second-order spin-orbit coupling of the model
were originally adjusted to reproduce Feshbach resonances due
to zero-energy bound states of $d$ and $g$ symmetry, the
positions of crossings between $s$, $d$, and $g$ levels tend to
be accurate to within the model uncertainties, which are on the
order of 0.05 mT or less~\cite{Leo2000,Chin:cs2-fesh:2004}. On
the other hand, the levels of $l$ symmetry, corresponding to
$L=8$, are off by up to 0.5 mT, a much larger amount. One
plausible reason for this has to do with the large rotational
energy of the $6l(M_F)$ levels that cross the $6g(6)$ level.
The $6g(6)$ level has the vibrational character of the second
$6s(6)$ vibrational level below the lowest separated-atom
limit, with about 110 MHz of $l=4$ rotational energy added. The
crossing $6l(M_F)$ levels, by contrast, have the  vibrational
character of the third vibrational level below the limit, with
about 740 MHz of rotational energy added to bring them near
threshold. More deeply bound  levels with more rotational
energy can have larger errors due to deficiencies in the model
potentials. An error of only a few parts per 1000 in the
rotational energy can lead to a 0.5 mT error in the crossing
positions for $6l(M_F)$ levels.

Additional information is contained in the
coupling strengths that govern the closest approach $2V$
between levels at avoided crossings. In the calculations, this
quantity is determined largely by the second-order spin-orbit
contribution to $\hat V^\mathrm{d}$. This is a relatively
poorly determined parameter in our model and is uncertain to
about 15\%~\cite{Leo2000}.

Several different experimental methods have been used to
determine coupling strengths. Mark {\em et al.}
\cite{Mark2007b} used a method based on St\"uckelberg
interferometry, which gives precise measurements of the energy
difference between the two states. Mark {\em et al.}
\cite{Mark2007a} used a different method based on integrating
magnetic moment values. This gives absolute energies for the
two states (rather than just the difference between them) but
is now believed to overestimate the coupling strengths in some
cases~\cite{Ferlaino:priv:2008}, especially for crossings
between states with very different magnetic moments. Some
crossing strengths were also estimated from a Landau-Zener
approach. Lastly, Ferlaino {\em et al.}
\cite{Ferlaino:wiggle:2008} have used a method in which
transitions are induced by modulating the magnetic field
\cite{Thompson:magres:2005}. This is the most precise of the
different methods.

\begin{table}[tbh]
\caption{Comparison of results from the best propagator
calculation with the experimental results for selected level
crossings with the 6g(6) state. The columns label the symmetry
of the crossing state, the origin of the value,
the $L$ functions in the basis set used for the calculation,
and the position $B_0$, energy $E/h$, and the energy splitting
$2V/h$ for each crossing.  The lines labeled ``Exp'' show the
experimental values. } \label{tab:experiment}
\begin{tabular}{crclll}
\hline\hline
State & method & basis  & $B_0$ [mT] & $E/h$ [MHz] & $2V/h$ [kHz] \\
\hline

6l(3)  & propagator & $gil$    & 0.9181    & -5.135   &  21.2      \\
       & Exp$^a$  &       & 1.122(2)  &          &  32(6)     \\
       & Exp$^b$  &       & 1.1339(1) &          &  28(2)     \\
6l(4)  & propagator & $gil$    & 1.2715    & -5.088   &  43.6      \\
       & Exp$^a$  &       & 1.550(3)  &          & 128(26)    \\
4g(4)  & propagator & $dgi$    & 1.375     & -5.11    & 277.       \\
       & Exp$^a$  &       & 1.329(4)  &          & 328(60)    \\
       & Exp$^c$  &       & 1.357(1)  &          & 291.4(8)   \\
6s(6)  & propagator & $sdgi$   & 1.8664    & -5.079   &  44.8      \\
       & Exp$^c$  &       & 1.8651(3) &          &  58(17)    \\
6l(5)  & propagator & $gil$    & 2.0089    & -5.056   &  66.6      \\
       & Exp$^a$  &       & 2.53(1)   &          & 126(44)    \\
4d(4)  & propagator & $sdgi$   & 4.4766    & -4.885   &  57.6      \\
       & Exp$^a$  &       & 4.515(4)  &          & 240(42)    \\
       & Exp$^c$  &       & 4.5106(3) &          &  78(9)     \\
\hline\hline
\end{tabular}\\
a. Reference \cite{Mark2007a}\\
b. Reference \cite{Mark2007b}\\
c. Reference \cite{Ferlaino:wiggle:2008}.
\end{table}

The crossing strengths for various different levels crossing
the $6g(6)$ level near 5 MHz are compared with the available
experimental values in Table~\ref{tab:experiment}. The most
reliable experimental results are those from
St\"uckelberg oscillations \cite{Mark2007b} and magnetic field
modulation \cite{Ferlaino:priv:2008} for the $6l(3)$, $4g(4)$,
$6s(6)$ and $4d(4)$ levels. The $6l(3)$ and $6s(6)$ levels are
indirectly coupled to $6g(6)$, and for both these the
calculated crossing strength is about 25\% lower than the best
experimental value. The $4g(4)$ and $4d(4)$ levels are directly
coupled to $6g(6)$; for the $4g(4)$ level the calculated
crossing strength is about 5\% lower than experiment, while for
the $6s(6)$ level the discrepancy is larger but is within the
experimental error bars. This suggests that the strength of the
coupling term $V^{\rm d}(R)$ is underestimated but within the error
range of Leo {\it et al.}~\cite{Leo2000}.

Some of the other crossings in Table~\ref{tab:experiment} show
larger differences between experiment and theory, but in all
these cases the experimental value was obtained using the less
reliable magnetic moment method. The possible experimental
errors for the magnetic moment approach are illustrated by the
$4d(4)$ crossing, where it gives a crossing strength a factor
of 3 larger than the more accurate magnetic field modulation
method.  It would be very interesting to remeasure the
$6l(4)$, $6l(5)$ and other crossings in order to establish
whether there is a consistent relative error between experiment
and theory.

Errors in the level positions can result from
deficiencies in either the long-range or the short-range part
of the model potentials. As discussed above, there are
remaining discrepancies in level positions of up to 0.05 mT for
$s$, $d$, and $g$ levels, and up to 0.5 mT for $l$ levels.
Further improvements in the potential model are thus needed for
this important prototype system. This is particularly important
for predicting the resonances and crossings in the 80 mT
region, where interesting Efimov physics is
predicted~\cite{Lee2007} and even greater sensitivity to model
errors is expected. A major advantage of the propagator method
introduced here is that it is inexpensive enough to be used to
determine model parameters by least-squares fitting to level
energies and locations and strengths of level crossings.

\section{Conclusions}

We have presented a new computational method for calculating
bound states of molecules such as Cs$_2$. The method
is based on solving a set of coupled differential equations by
propagation, without relying on a basis set for the interatomic
coordinate. This is much more efficient than using a radial
basis set and allows the use of much larger basis sets of spin
functions. It also eliminates problems with calculating bound
states very near to dissociation, because the propagation can
be extended to very large separations at very little expense.
The new method makes it possible for the first time to carry
out fully converged calculations on bound states of Cs$_2$,
including anisotropic couplings due to spin-spin and
second-order spin-orbit interactions, and to
characterize avoided crossings between pairs of levels.

We have compared the results of converged calculations using
the current best Cs$_2$ model potentials with experimental
measurements on the near-dissociation states of Cs$_2$ in a
magnetic field. The model generally performs well for $s$, $d$
and $g$ states (with $L=0$, 2 and 4), though even there there
are quantitative discrepancies of up to 0.05 mT in the magnetic
fields at which levels cross. The discrepancies are much larger
(0.5 mT) for $l$ states ($L=8$).  The strengths of
the avoided crossings also appear to be systematically
underestimated by the current model. These discrepancies should
in future allow the development of improved models for the
potential curves and couplings in the Cs$_2$ dimer. Such model
improvement is both desirable and possible, not only for
near-threshold levels but also to provide an improved
representation of more deeply bound states such as those
measured by Vanhaecke {\em et al.}~\cite{Vanhaecke2004}.
High-quality models are also important for proposals to use
precision measurements on Cs$2$ for fundamental physics
studies~\cite{Chin2006,DeMille2008}.

\section{Acknowledgements}
P.S. Julienne acknowledges the Office of Naval Research for
partial support. J. M. Hutson is grateful to EPSRC for support
under the ESF EUROCORES Programme EuroQUAM.

\appendix\section{Matrix elements\label{app}}
\begin{widetext}

In the decoupled basis set (\ref{eqbasdecoup}), the matrix
elements of the isotropic potential operator $\hat V^{\rm
c}(R)$ between primitive (unsymmetrized) basis functions are
\begin{eqnarray}
\langle s_1 m_{s1} i_1 m_{i1} s_2 m_{s2} i_2 m_{i2} L M_L | \hat
V^{\rm c}(R) | s_1 m_{s1}^\prime i_1 m_{i1}^\prime s_2
m_{s2}^\prime i_2 m_{i2}^\prime L^\prime M_L^\prime \rangle =
\delta_{L L^\prime} \delta_{M_L M_L^\prime}
\delta_{m_{i1} m_{i1}^\prime} \delta_{m_{i2} m_{i2}^\prime}
\nonumber\\\sum_{S} V_S(R)
(-1)^{2s_1-2s_2+m_{s1}+m_{s2}+m_{s1}^\prime+m_{s2}^\prime} (2S+1)
\left(\matrix{ s_1 & s_2 & S \cr m_{s1} & m_{s2} & -m_{s1}-m_{s2}}
\right)
\left(\matrix{ s_1 & s_2 & S \cr m_{s1}^\prime & m_{s2}^\prime &
-m_{s1}^\prime-m_{s2}^\prime} \right).
\end{eqnarray}
The corresponding matrix elements of the spin-spin operator are
\begin{eqnarray}
\langle s_1 m_{s1} i_1 m_{i1} s_2 m_{s2} i_2 m_{i2} L M_L | \hat
V^{\rm d}(R) | s_1 m_{s1}^\prime i_1 m_{i1}^\prime s_2
m_{s2}^\prime i_2 m_{i2}^\prime L^\prime M_L^\prime \rangle =
\delta_{m_{i1} m_{i1}^\prime} \delta_{m_{i2} m_{i2}^\prime}
\lambda(R)
\nonumber\\
(-1)^{s_1+s_2-m_{s1}-m_{s2}-M_L}
\left[s_1(s_1+1)(2s_1+1) s_2(s_2+1)(2s_2+1)
(2L+1)(2L^\prime+1)\right]^{1/2}
\left(\matrix{ L & 2 & L^\prime \cr 0 & 0 & 0} \right)
\nonumber\\ \sum_{q_1 q_2}
\left(\matrix{ L & 2 & L^\prime \cr -M_L & -q_1-q_2  & M_L^\prime}
\right)
\left(\matrix{ 1 & 1 & 2 \cr q_1 & q_2 & -q_1-q_2} \right)
\left(\matrix{ s_1 & 1 & s_1 \cr -m_{s1} & q_1 & m_{s1}^\prime}
\right)
\left(\matrix{ s_2 & 1 & s_2 \cr -m_{s2} & q_2 & m_{s2}^\prime}
\right),
\end{eqnarray}
where for any individual matrix element the sums over $q_1$ and
$q_2$ collapse because of the selection rules imposed by the last
two 3-$j$ symbols. The matrix elements of the atomic nuclear spin
operators are particularly simple in this basis set,
\begin{eqnarray}
\langle s_1 m_{s1} i_1 m_{i1} s_2 m_{s2} i_2 m_{i2} L M_L |
\hat\imath_1 \cdot \hat s_1 | s_1 m_{s1}^\prime i_1 m_{i1}^\prime
s_2 m_{s2}^\prime i_2 m_{i2}^\prime L^\prime M_L^\prime \rangle =
\nonumber\\ \delta_{LL^\prime} \delta_{m_{s2} m_{s2}^\prime}
\delta_{m_{i2} m_{i2}^\prime}
\langle s_1 m_{s1} i_1 m_{i1}  | \hat\imath_1 \cdot \hat s_1 | s_1
m_{s1}^\prime i_1 m_{i1}^\prime \rangle
\end{eqnarray}
where
\begin{eqnarray}
\langle s_1 m_{s1} i_1 m_{i1}  | \hat\imath_1 \cdot \hat s_1 | s_1
m_{s1} i_1 m_{i1} \rangle &=& m_{i1} m_{s1}; \\
\langle s_1 m_{s1} i_1 m_{i1}  | \hat\imath_1 \cdot \hat s_1 | s_1
m_{s1}\pm1 i_1 m_{i1}\mp1 \rangle &=&
\left[s_1(s_1+1)-m_{s1}(m_{s1}\pm1)\right]^{1/2}
\left[i_1(i_1+1)-m_{i1}(m_{i1}\mp1)\right]^{1/2},
\end{eqnarray}
and similarly for $\hat\imath_2 \cdot \hat s_2 $. The matrix
elements of $\hat L^2$ are simply
\begin{eqnarray}
\langle s_1 m_{s1} i_1 m_{i1} s_2 m_{s2} i_2 m_{i2} L M_L | \hat
L^2 | s_1 m_{s1}^\prime i_1 m_{i1}^\prime s_2 m_{s2}^\prime i_2
m_{i2}^\prime L^\prime M_L^\prime \rangle =
\nonumber\\ \delta_{LL^\prime} \delta_{M_LM_L^\prime}
\delta_{m_{s1} m_{s1}^\prime} \delta_{m_{i1} m_{i1}^\prime}
\delta_{m_{s2} m_{s2}^\prime} \delta_{m_{i2} m_{i2}^\prime}
L(L+1) \,,
\end{eqnarray}
and those of the Zeeman operator are
\begin{eqnarray}
\langle s_1 m_{s1} i_1 m_{i1} s_2 m_{s2} i_2 m_{i2} L M_L | \hat
g_e \mu_{\rm B}
B \, \hat s_{zj} + g_n \mu_{\rm B} B \, \hat \imath_{zj}
| s_1 m_{s1}^\prime i_1 m_{i1}^\prime s_2 m_{s2}^\prime i_2
m_{i2}^\prime L^\prime M_L^\prime \rangle =
\nonumber\\ \delta_{LL^\prime} \delta_{M_LM_L^\prime}
\delta_{m_{s1} m_{s1}^\prime} \delta_{m_{i1} m_{i1}^\prime}
\delta_{m_{s2} m_{s2}^\prime} \delta_{m_{i2} m_{i2}^\prime}
(g_e \mu_{\rm B} B \, \hat m_{sj} + g_n \mu_{\rm B} B \, m_{ij}).
\end{eqnarray}

All the calculations in the present paper used basis functions
symmetrized for exchange of two identical particles with
$s_1=s_2=s$ and $i_1=i_2=i$. For $m_{s1}=m_{s2}$ or
$m_{i1}=m_{i2}$ the symmetrized functions are identical to the
unsymmetrized ones, except that only even $L$ is allowed for
bosons and only odd $L$ for fermions. For $m_{s1}\ne m_{s2}$ or
$m_{i1}\ne m_{i2}$, the symmetrized functions are
\begin{equation}
\left[ | s m_{s1} i m_{i1} s m_{s2} i m_{i2} L M_L
\rangle \pm (-1)^L | s m_{s2} i m_{i2} s m_{s1} i m_{i1} L M_L
\rangle \right] /\sqrt{2} \,,
\end{equation}
with the $+$ sign for bosons and the $-$ sign for fermions.

The Hamiltonian in the basis set,
Eq.~(\ref{eqbascoup}), used in the DVR calculations can be
derived from the Hamiltonian in the uncoupled basis by
performing a unitary transformation, namely,  the
transformation $|\alpha_j m_{fj}\rangle$ to $|s_j m_{sj}\rangle
| i_j m_{ij}\rangle$ for each of the two atoms ($j$=1 or 2).
The transformation depends on the magnetic field strength
\cite{Breit1931}. In practice, the eigenvectors for the monomer
$h_j$ must be evaluated. As $m_{fj}$ is conserved at most a
2$\times$2 matrix needs to be diagonalized. Bose/Fermi
symmetrization is ensured by
\[
        [
             |\alpha_1 m_{f1}\alpha_2 m_{f2},LM_L\rangle
             \pm (-1)^L |\alpha_2 m_{f2}\alpha_1 m_{f1},LM_L\rangle
        ]/\sqrt{2}
\]
when $\alpha_1 \neq \alpha_2$ or $m_{f1}\neq m_{f2}$.  The
state with $\alpha_1 = \alpha_2$ and $m_{f1}= m_{f2}$ exists
only for even (odd) $L$ for bosonic (fermionic) atoms
respectively.
\end{widetext}

\bibliography{Cs2-bound_8}

\end{document}